\documentstyle[11pt,paspconf,epsf]{article}
\begin{document}

\title{WIYN Open Cluster Study: Lithium in Cool Dwarfs of the M35 Open Cluster}

\author{David Barrado y Navascu\'es}
\affil{Max-Planck-Institut f\"ur Astronomie, Heidelberg,  Germany}

\author{Constantine P. Deliyannis\altaffilmark{1}}
\affil{Astronomy Department, Indiana University, USA}

\author{John R. Stauffer}
\affil{Harvard--Smithsonian Center for Astrophysics, Cambridge, USA}

\altaffiltext{1}{Visiting WIYN Observer. The WIYN Observatory is a 
joint facility of the University of
    Wisconsin-Madison, Indiana University, Yale University, and the
    National Optical Astronomy Observatories.}

\begin{abstract}
We have obtained high resolution spectra of $\sim$40 members of M35,
 determined the  
Lithium-T$_{\rm{eff}}$ morphology and the distribution of the rotational
velocity for G and K stars, and compared them
to those of the Pleiades and other well-known open clusters.
\end{abstract}

\section{The cluster background}

M35 (NGC~2168) is a very interesting open cluster for several reasons. 
It is  relatively nearby, with (m-M)$_0$=9.7 (Vidal 1973). It is very well
 populated, with a total mass estimated between 1000 and 3000 M$_\odot$. 
While M35 is at fairly low galactic latitude
({\it l}$^{\sc \,II}$=186.58, {\it b}$^{\sc \,II}$=2.19),
the cluster is rich enough so that  it is easy to 
separate the field stars and the cluster population.
At the same time, the reddening is not very high 
 --E(B--V)=0.17, (Vidal 1973).
Finally, it has been considered a coeval cluster to the Pleiades, since
their turn-off points are located at the similar position  on the 
color-magnitude diagram. 
The traditional age of the Pleiades is 70-100 Myr. 

\begin{figure}	
\vspace{-1.0cm}
\plotone{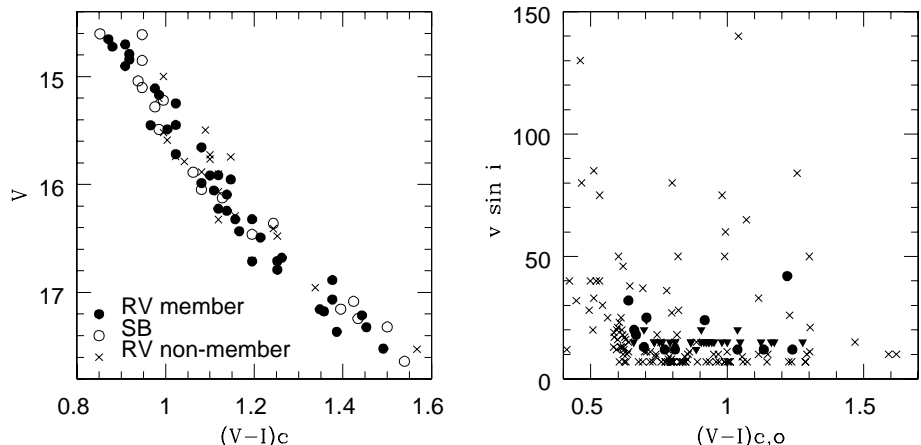}
\vspace{-7.2cm}
\caption{{\bf a} Color-Magnitude diagram for our M35 candidate members.
{\bf b}  Projected rotation velocities against the unreddened
(V-I)$_C$ color for M35 (solid circles and triangles --upper limits--.
Only RV members included) and the Pleiades (crosses).} 
\end{figure}

\section{The selection of  members.}

We selected  cluster member candidates   
based on the position of the  [V,(V--I)$_C$] color-magnitude diagram
(see Barrado y Navascu\'es et al. 1999).
The   magnitude interval  was 14.5 $\le$ V  $\le$ 17.5,
whereas the color range was   0.8 $\le$ (V-I)$_{\rm c}$ $\le$ 1.6. 
Since M35 is very close to the galactic plane, we expect 
some contamination by field stars. In fact, there is not
a clear gap between the locus of the M35 MS and field stars.
This smooth transition indicates that a list of photometrically selected
 candidates of M35 should  contain  a significant
number of spurious members.
Subsequently, our member candidates were observed spectroscopically
using WIYN/HYDRA, a multifiber spectrograph, capable of
observing 97 targets simultaneously at R$\sim$20,000 (as measured in the
comparison lamps, $\sim$2 pixels).
We took 6 individual exposures of $\sim$2 hours each, over 2 different
nights. After processing and extracting the individual
spectra, we added up all of them to create high quality spectra. 
Final signal-to-noise ratios range from 40 to 160 per pixel.
 We measured radial velocities (RV) in the final  as well 
as in the individual spectra, using several iron lines
around the lithium doublet (LiI6707.8 \AA). Using the
measured RV, we catalogued our photometric candidate members
in three different groups:
\begin{itemize}
\item Probable members: This group includes the stars with 
non-variable RV, which value is close to the average of the cluster
(38 stars).
\item Probable short-period binaries, possible members:
 A significant fraction 
of our sample presents variability in the RV. Since we only have 
6 points, we cannot derive the value of the center of gravity, and we 
are unable to establish if these stars belong to the open cluster
(19 stars).
\item Probable field stars (non-members)
or long-period binaries  (possible members):
 Those stars with an apparent  fixed value of RV, very 
different of the value of the cluster (19 stars).
\end{itemize}
Figure 1a shows a color-magnitude diagram with 
 all the possible candidates identified with our
optical survey which were observed  spectroscopically.
Probable members are depicted with solid circles, whereas
 SB and non-members appear as 
open circles and crosses, respectively.
Note that the  short-period SB
 stars represented a third of the probable/possible
members of M35. By comparison, Praesepe has $\sim$40\% of SB.
Therefore, we  expect some of these   SB stars to be
cluster members.


\section{The relation Lithium--Teff--Rotation}

\begin{figure}	
\vspace{-1.0cm}
\plotone{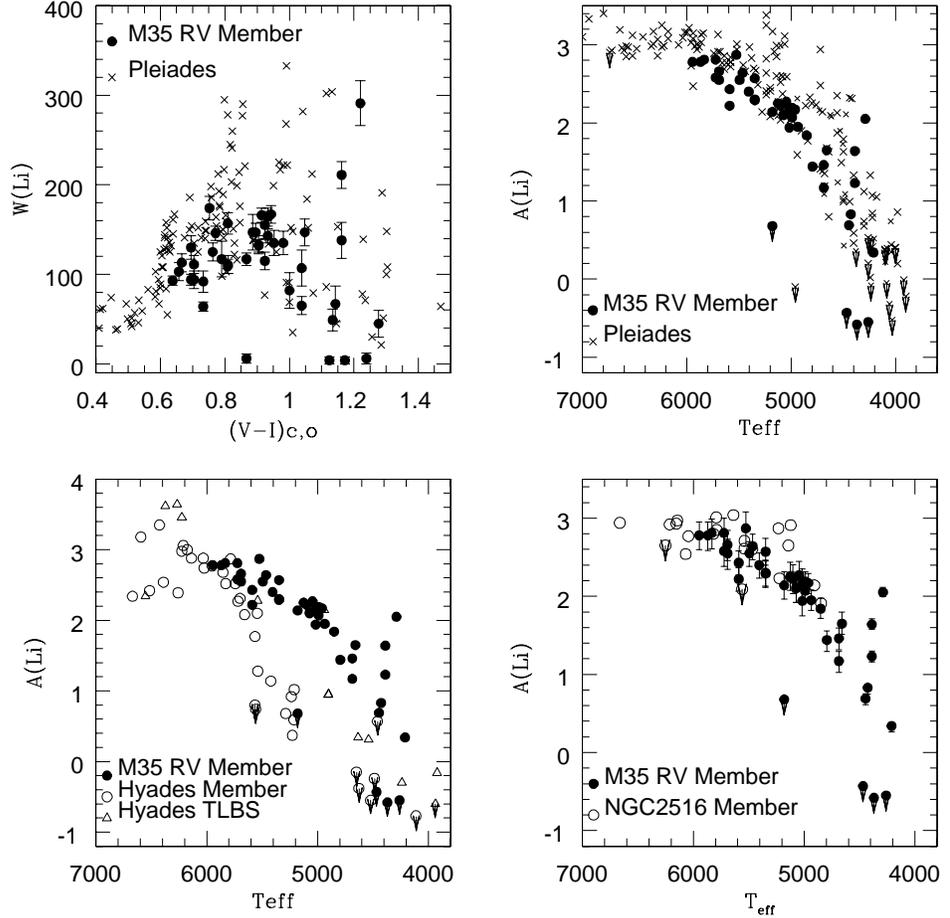}
\vspace{-0.8cm}
  \caption{{\bf a} Li equivalent widths. Other panels, Li-Teff plane:
 {\bf b} Comparison with 
the Pleiades, {\bf c} the Hyades, {\bf d} NGC2516.}
\end{figure}

We have measured the projected rotation velocities 
and compare with their distribution in the Pleiades (Figure 1b)
--Pleiades data from  Soderblom et al. (1993).
Contrary to what happens in the Pleiades, 
 M35 lacks fast rotators in this color range. Since stars spin-down
with age, M35 could be older than the Pleiades. However, the 
difference in the distribution of the rotational velocity also could be due 
to the initial conditions.
 Li abundances --A(Li)-- were derived from  the LiI6708\AA{ }
equivalent widths (W), taking into account the FeI6707.4\AA{ } line, and 
using curves of growth from Soderblom et al. (1993).  Comparison between
M35 and the Pleiades  shows that although the 
shape of the dependence of the W(Li) with color is quite similar, the
M35 values are systematically
 smaller and the scatter for a given color is also  reduced
(Figure 2a).
The Li-Teff plane is depicted in Figures 2b,c,d. 
 Panel b shows the Pleiades,
Panel c displays  the Hyades (see Barrado y Navascu\'es \& Stauffer 1996
 and references therein),
 and Panel d contains NGC2516 data (Jeffries et al. 1998). 
Three main studies have been carried out concerning the relation between
the spread in the Li abundance and rotation-activity in the 
Pleiades open cluster (Soderblom et al. 1993;
 Garc\'{\i}a-L\'opez et al. 1994; Jones et al. 1996).
 All these three works concluded that
Pleiades stars present an intrinsic scatter in their
abundances for the same color and connected the
differences with stellar activity and rotation
 (the higher the activity, the
faster the rotation and the Li abundance for a given color).
 In the case of M35, we do not see a dramatic scatter in the Li abundance,
 at least in the range 6000$>$Teff$>$4600. Below that temperature, 
a more important scatter  is  present. However, only two stars have
 an abundance  notably higher than the average trend. 
In this color range, a connection between rotation and
 Li might be present, since the only obvious fast rotator (42 km/s) is
 clearly  a Li--rich M35 star. 
Finally, the average M35 Li  abundance for a given color
is systematically smaller than the average Pleiades Li abundance.
The comparisons with other open clusters, such as the Hyades and
NGC2516 (Panels c and d) indicate that, in fact, M35 could be
as old as this last one, based on the morphology of the
dependence  of Li with effective temperature, if  we assume
 that age is the single most important parameter
for Li depletion in cool dwarfs (see Deliyannis 1999). 


\section{Conclusions.}

 We have detected a significant number of late spectral
type members of the M35 open cluster based on optical
photometry and high resolution spectroscopy and
 derived Li abundances for our
M35 members. We have compared their distribution with Pleiades stars.
Contrary to what happens in the Pleiades, 
M35 does not show a significant scatter 
in the Li-Teff plane for dG and early dK stars.
The derived abundances are systematically smaller.
On the other hand,  the distribution of vsini are also different 
in both clusters,
with M35 lacking  fast rotators; a connection between rotation and 
Li abundance might be present in late dK members of M35.
Therefore, M35 seems to be older than
 the Pleiades, perhaps as old as NGC 2516 (140 Myr), which agrees
with resent results (von Hippel et al. 1999; Sung \& Bessel 1999).

\acknowledgments
DByN thanks the IAC (Spain) 
and the DFG (Germany) for their fellowship,
and the support by the European Union.

\end{document}